\documentclass[12pt]{article}

\usepackage{amssymb,amsmath}

\usepackage{graphicx}
\DeclareGraphicsExtensions{.pdf,.png,.jpg}

 \catcode`\@=11 \@addtoreset{equation}{section}\catcode`\@=12

\newcommand{\R}{ {\mathbb R} }
\newcommand{\eps}{ \varepsilon }

\begin{document}

 \begin{center}

 \large \bf Stable  exponential cosmological type solutions with three 
 factor spaces in EGB  model  with a $\Lambda$-term 
  \end{center}

 \vspace{0.3truecm}

 \begin{center}

   K. K. Ernazarov$^{1}$,  V. D. Ivashchuk$^{1,2}$

\vspace{0.3truecm}

  \it $^{1}$ 
    Institute of Gravitation and Cosmology, \\
    Peoples' Friendship University of Russia (RUDN University), \\
    6 Miklukho-Maklaya Street,  Moscow, 117198, Russian Federation, \\ 
 
  \it $^{2}$ Center for Gravitation and Fundamental Metrology,  VNIIMS, \\
  46 Ozyornaya Street, Moscow, 119361,  Russian Federation.

\end{center}

\begin{abstract}

We study a $D$-dimensional  Einstein-Gauss-Bonnet model which includes  the Gauss-Bonnet term,  the cosmological term $\Lambda$ and two  non-zero constants: $\alpha_1$ and $\alpha_2$. Under imposing the metric to be diagonal one, we  find   cosmological type  solutions with  exponential dependence of three scale factors  in a variable $u$, governed by three non-coinciding Hubble-like parameters: $H \neq 0$, $h_1$ and $h_2$, obeying  $m H + k_1 h_1 + k_2 h_2 \neq 0$,  corresponding to factor spaces of dimensions $m > 1$, $k_1 > 1$ and $k_2 > 1$, respectively,  and depending upon sign parameter   $\varepsilon = \pm 1$, where $\varepsilon = 1$ corresponds to cosmological case  and $\varepsilon = - 1$ - to static one). We deal with two cases: i) $m  <  k_1 < k_2$ and ii) $1< k_1 = k_2 = k$, $k \neq m$. We show that in both cases the solutions exist if $\varepsilon \alpha = \varepsilon \alpha_2 / \alpha_1 > 0$ and $\alpha \Lambda > 0$ satisfies certain (upper and lower) bounds. The solutions are defined up to solutions of certain polynomial master equation  of order four (or less)  which may be solved in radicals.  In case ii) explicit solutions are presented. In both cases  we single out  stable and non-stable solutions as $u \to \pm \infty$. The case $H = 0$ is also considered.

\end{abstract}

  {\bf Keywords:} Gauss-Bonnet, dark energy, stability

\section{Introduction}

In this semi-review article, which generalizes our previous work \cite{ErIv-20}, we deal with  the so-called  Einstein-Gauss-Bonnet (EGB) gravitational model in dimensions  $D > 7$, which contains   Gauss-Bonnet term and cosmological term $\Lambda$. The model  also includes two non-zero constants: $\alpha_1$ and $\alpha_2$, corresponding to Einstein and Gauss-Bonnet terms, respectively. It is well-known that the equations of motion for this model are of the second order (as it appears in General Relativity). The so-called Gauss-Bonnet term  has appeared in (super)string theory  as a second order correction in curvature  to the  effective (super)string effective action   \cite{Zwiebach,FrTs2}. 

At present, EGB gravitational model, e.g. with cosmological term,   and  its modifications   \cite{Ishihara}-\cite{NOO-19} ,  are under intensive  studyies in  cosmology and astrophysics, aimed at solution of dark energy problem, i. e. possible explanation  of  accelerating  expansion of the Universe, which follows from supernovae (type Ia) observational data \cite{Riess,Perl}, and search of possible local manifestation of dark energy (related to black holes, wormholes etc).

In this article  we  start with the so-called cosmological type solutions with ``diagonal'' metric  
$ds^2 = - \eps (d u)^2  +  \sum_{i=1}^{n} \eps_i a_i^2(u)  (dy^i)^2$,  governed by $n >3$ scale factors ($D = n+1$, 
$\eps = \pm 1$, $\eps_i = \pm 1$) depending upon one variable $u$, which is the synchronous time variable for cosmological case, when $\eps = \eps_i = 1$. For the case  $\eps = -1$ and $\eps_1 = -1$,  $\eps_j = 1$ ($j > 1$) we get static configurations described by space-like variable (coordinate) $u$ and time-like coordinate  $y^1$. In cosmological case the equations of motion are governed by an effective Lagrangian which contains $2$-metric (or minisupermetric) $G_{ij}$ and finslerian metric $G_{ijkl}$, see Refs. \cite{Iv-09,Iv-10}  for $\Lambda = 0$ and  Ref.  \cite{IvKob-18mm} for $\Lambda \neq 0$.  

Here we consider the cosmological type solutions  with exponential dependence of scale factors (upon $u$-variable)  and obtain a class of  solutions with  three scale factors, governed by three non-coinciding Hubble-like parameters: $H$, $h_1$ and $h_2$, corresponding to factor spaces of dimensions $m > 1$, $k_1 > 1$ and $k_2 > 1$, respectively ($D = 1 + m + k_1 + k_2$). Here we impose the  following restriction $S_1 = m H + k_1 h_1 + k_2 h_2 \neq 0$,   excluding the solutions with constant volume factor and addressing us  to a classification theorem which tells us   that  for  generic anisotropic exponential solutions with Hubble-like parameters $h_1, \dots, h_n$  obeying  $S_1 = \sum_{i=1}^n h_i \neq 0$  the number of different (real) numbers among  $h_1, \dots, h_n$  may be $1$, or $2$, or $3$ \cite{Ivas-16}. 
The main goal of this paper is to extend the results of Ref. \cite{ErIv-20} to a class of  cosmological type solutions, which include static ones (with $\eps = - 1$). 

Here, as in Ref. \cite{ ErIv-20}   we consider  without loss of generality two cases: i) $m <  k_1 < k_2$ and ii) $1< k_1 = k_2 = k$, $k \neq m$. ( In the case $m =  k_1 = k_2$ the solutions are absent due to our restrictions.) 
For $H \neq 0$ in both cases the solutions  exist  only if $\alpha \eps = \eps \alpha_2 / \alpha_1 > 0$, $ \Lambda \eps > 0$ and  multidimensional  cosmological term $\Lambda$ obeys the bounds:  $0 < \lambda_{-}(m,k_1,k_2) \leq  \Lambda \alpha \leq  \lambda_{+}(m,k_1,k_2)$.  For  $H =0$ the solutions
exist only when $\alpha \eps > 0$, $ \Lambda \eps > 0$, $k_1 \neq k_2$ and   $\Lambda \alpha =  \lambda_{\infty }(k_1,k_2) >0$. We note that here, as in Ref. \cite{ErIv-20} we use the Chirkov-Pavluchenko-Toporensky scheme of  reduction  of the set of polynomial equations  \cite{ChPavTop1}. As in Ref. \cite{ ErIv-20}   we reduce the problem
in generic $H \neq 0$ case to solutions of a single polynomial master equation  of fourth order or less, which may be solved in radicals for all $m > 1$, $k_1 > 1$ and $k_2 > 1$.
In the case ii) $1< k_1 = k_2 = k$, $k \neq m$  ($H \neq 0$) the solutions for Hubble-like parameters are found explicitly (see Section 4).  
 
 We also study (in Section 5) the stability of the  solutions for $u \to \pm \infty$ in a class of cosmological type solutions with diagonal metrics by using an extension of results of  Refs. \cite{Ivas-16, ErIv-20} (see also  approach of Ref. \cite{Pavl-15}) and  single out  the subclasses of stable/non-stable solutions.

We note that the exponential cosmological type solutions with  two non-coinciding Hubble-like parameters   $H \neq 0$ and $h$ obeying $S_1 = m H + l h_1 \neq 0$ with $m > 2$, $l>2$  were studied earlier in Ref. \cite{ I-20}. In that case  there were two sets of solutions obeying: a) $\eps \alpha > 0$, $ \alpha  \Lambda < \lambda_{+}(m,l)$ and  b) $\eps  \alpha < 0$, $ \alpha \Lambda < - \lambda_{-}(m,l)$, where $\lambda_{\pm}(m,l) > 0$ and $\eps = \pm 1$. 

It should be noted that recently EGB models were used for constricting certain 4-dimensional gravitational models (so-called 4DEGB theories, e.g. belonging to Horndeski class) by using ideas of Glavan-Lin  rescaling \cite{GL}  and/or dimensional reductions. These 4D modified models of gravity are (at the moment) under intensive study and have numerous applications  in gravitational physics and cosmology, for a review see Ref. \cite{FCCM}.

\section{The cosmological model}

We start with the model governed by the action
\begin{equation}
  S =  \int_{M} d^{D}z \sqrt{|g|} \{ \alpha_1 (R[g] - 2 \Lambda) +
              \alpha_2 {\cal L}_2[g] \}.
 \label{2.0}
\end{equation}
Here, $g = g_{MN} dz^{M} \otimes dz^{N}$ is the metric  on
a manifold $M$ (${\dim M} = D$), $|g| = |\det (g_{MN})|$, $\Lambda$ is
the cosmological term, $R[g]$ is scalar curvature,
$${\cal L}_2[g] = R_{MNPQ} R^{MNPQ} - 4 R_{MN} R^{MN} +R^2$$
is the  Gauss-Bonnet term and  $\alpha_1$, $\alpha_2$ are
certain nonzero constants.

Our choice of the manifold is following
\begin{equation}
   M = \R  \times   M_1 \times \ldots \times M_n. 
   \label{2.1}
\end{equation}
In what follows we deal with the metric
\begin{equation}
   g= -  \eps d u \otimes d u  +
      \sum_{i=1}^{n} B_i \eps_i e^{2v^i u} dy^i \otimes dy^i.
  \label{2.2}
\end{equation}
Here   $B_i > 0$ are arbitrary constants, $\eps = \pm 1$, $\eps_i = \pm 1$, $i = 1, \dots, n$
($n > 3$) and $M_1, \dots,  M_n$  are chosen to be 1-dimensional manifolds 
(either non-compact ($\R$) or compact ($S^1$) ones).
The cosmological case ($\eps =\eps_i =  1$) was considered in detail 
in Ref. \cite{ErIv-20}. The case $\eps = - 1$ may  describe certain static configurations.

 The action (\ref{2.0})  with the ansatz for the metric (\ref{2.0}) imposed 
 gives rise to the  equations of motion which 
 are of polynomial type \cite{ErIvKob-16}
\begin{eqnarray}
  E = G_{ij} v^i v^j + 2 \Lambda \eps
  - \alpha \eps  G_{ijkl} v^i v^j v^k v^l = 0,  \label{2.3} \\
   Y_i =  \left[ 2   G_{ij} v^j
    - \frac{4}{3} \alpha \eps  G_{ijkl}  v^j v^k v^l \right] \sum_{i=1}^n v^i 
    - \frac{2}{3}   G_{ij} v^i v^j  +  \frac{8}{3} \Lambda \eps = 0,
   \label{2.4}
\end{eqnarray}
$i = 1,\ldots, n$.    Here we denote $\alpha = \alpha_2/\alpha_1$ and
\begin{equation}
G_{ij} = \delta_{ij} -1, \qquad   G_{ijkl}  = G_{ij} G_{ik} G_{il} G_{jk} G_{jl} G_{kl},
\label{2.4G}
\end{equation}
\cite{Iv-09,Iv-10}. 
For $n > 3$ we have a set of polynomial  equations of order $4$.

For the case $n > 3$,  $\Lambda =0$ and $\alpha \eps  < 0$ the set of equations (\ref{2.3}) 
and (\ref{2.4}) has a trivial (isotropic) solution: $v^1 = \cdots = v^n = H$  \cite{Iv-09,Iv-10},
which was generalized in Ref. \cite{ChPavTop} to the case $\Lambda \neq 0$.

In Ref. \cite{Iv-09,Iv-10} the following proposition was proved: there are no more than
three different  numbers among  $v^1,\dots ,v^n$ if $\Lambda =0$. This proposition 
was generalised in ref. \cite{Ivas-16} for  $\Lambda \neq 0$, when
the following condition is imposed $\sum_{i = 1}^{n} v^i \neq 0$.


In this paper  we study solutions to  equations (\ref{2.3}), 
(\ref{2.4}) by using the following ansatz:
\begin{equation}
  \label{3.1}
   v =(\overbrace{H, \ldots, H}^{m}, 
   \overbrace{h_1, \ldots, h_1}^{k_1}, \overbrace{h_2, \ldots, h_2}^{k_2}).
\end{equation}
Here $H$ is the Hubble-like parameter which corresponds  
to an $m$-dimensional factor space with inequality $m > 1$ imposed, while $h_1$ is the Hubble-like parameter 
which is related to an $k_1$-dimensional factor space with $k_1 > 1$ and 
$h_2$ is the Hubble-like parameter assigned to an $k_2$-dimensional factor space with $k_2 > 1$. 

 In what follows we add additional restrictions to our ansatz (\ref{3.1}):  
   \begin{equation}
     H \neq h_1, \quad  H \neq h_2, \quad 
     h_1 \neq h_2, \quad S_1 = m H + k_1 h_1 + k_2 h_2 \neq 0.
   \label{3.3}
   \end{equation}

 It was shown in Ref. \cite{ErIv-17-2} that the set of $(n + 1)$ polynomial equations  
 (\ref{2.3}), (\ref{2.4}) under ansatz  (\ref{3.1}) and restrictions (\ref{3.3}) 
 obeyed  are equivalent to a set  of polynomial equations 
     \begin{eqnarray}
          E =0,   \label{3.4E} \\
          Q =  - \frac{1}{2 \alpha \eps}, \label{3.4Q} \\
          L = H + h_1 + h_2 - S_1 = 0.  \label{3.4L}
     \end{eqnarray}
  which are of fourth, second and first orders, respectively.
 Here  $E$ is defined in (\ref{2.3}) and 
   \begin{equation}
        Q = Q_{h_1 h_2} =  S_1^2 - S_2 - 2 S_1 (h_1 + h_2) + 2 (h_1^2 + h_1 h_2 + h_2^2),
                   \label{3.5}
        \end{equation}
   where 
        \begin{equation}
       S_k = \sum_{i =1}^n (v^i)^k.
       \label{3.5a}
        \end{equation}   
  For more general prescription of scheme of reduction of polynomial equations of motion see 
  Ref. \cite{ChPavTop1} (the so-called Chirkov-Pavluchenko-Toporensky trick). 
 
 Relation (\ref{3.4Q}) is a special case of more general relations  \cite{ErIv-17-2}
  \begin{equation}
   Q_{h_i h_j} =  S_1^2 - S_2 - 2 S_1 (h_i + h_j) + 2 (h_i^2 + h_i h_j + h_j^2)
      = - \frac{1}{2 \alpha \eps}, \quad i \neq j,
       \label{3.5b}
   \end{equation}
 $i, j = 0,1,2$, with notation $h_0 = H$ used.       

Relation (\ref{3.3}) excludes the following case $H = h_1 = h_2 = 0$.  
In the main body of the paper we put
\begin{equation}
  \label{3.2a}
   H \neq 0. 
\end{equation}
          
      As in Ref. \cite{ErIv-20} we denote 
       \begin{equation}
         x_1 = h_1/H, \qquad     x_2 = h_2/H.      
                  \label{3.8}
       \end{equation}        
    In terms of dimensionless parameters the restrictions (\ref{3.3}) 
     may rewritten as following 
      \begin{equation}
        x_1 \neq 1, \quad  x_2 \neq 1, \quad  x_1 \neq x_2, 
        \quad  m  + k_1 x_1 + k_2 x_2 \neq 0.
      \label{3.3a}
      \end{equation}
        Equation  (\ref{3.4L}) is equivalent to the following one
   \begin{equation}
    m -1  + (k_1 - 1) x_1 + (k_2 -1) x_2 = 0.  
   \label{3.15}
   \end{equation} 
   
   In what follows we do not consider the case 
    \begin{equation}
      \label{3.1a}
      m = k_1 = k_2,
    \end{equation}
    which lead us to the empty set of solutions, 
    since we find for  $m = k_1 = k_2 >1 $ from restriction (\ref{3.3a}): 
    $1  +  x_1 + x_2 \neq 0$, while (\ref{3.15}) implies $1 + x_1 + x_2 = 0$.

    Due to (\ref{3.4Q}) and (\ref{3.5}) we obtain 
           
     \begin{equation}
          2 \alpha \eps {\cal P} H^2   = - 1,       \label{3.9}
      \end{equation} 
     where
     \begin{eqnarray}
     {\cal P}   = {\cal P} (x_1,x_2) 
     \nonumber \\    
     (m + k_1 x_1 + k_2 x_2 )^2  - (m + k_1 x_1^2 + k_2 x_2^2) 
     \nonumber \\
     - 2 (m + k_1 x_1 + k_2 x_2 )(x_1 + x_2) + 2 (x_1^2 + x_1 x_2 + x_2^2).
            \label{3.10}
     \end{eqnarray} 
 The relation (\ref{3.9}) is valid for  $ \alpha \eps {\cal P} < 0$. 
 It can be readily proved that \cite{ErIv-20} 
 \begin{equation}
         {\cal P} < 0       \label{3.10a}
 \end{equation}
 for $m > 1$, $k_1 > 1$, $k_2 > 1$. Indeed \cite{ErIv-20},
  \begin{eqnarray}
        {\cal P}    
        = 1 - m + (1 - k_1) x_1^2 +  (1 - k_2) x_2^2 < 0. 
                       \label{3.10b}
  \end{eqnarray} 
  It follows from (\ref{3.10a}) that
  \begin{equation}
       \alpha \eps  >  0.       \label{3.10c}
  \end{equation}
    
     The equation (\ref{3.4E}) reads  \cite{ErIv-20}
   \begin{eqnarray}
   2 \Lambda \eps = - G_{ij} v^i v^j +
     \alpha  \eps  G_{ijkl} v^i v^j v^k v^l 
                             \nonumber \\
     = H^2 V_1  +  \alpha \eps H^4 V_2, 
                         \label{3.11}    
   \end{eqnarray}  
 where 
  \begin{eqnarray}
   V_1 = V_1(x_1,x_2) 
    \nonumber \\
   =  - m - k_1 x_1^2 - k_2 x_2^2 + (m  + k_1 x_1 + k_2 x_2)^2 \label{3.12a} 
   \end{eqnarray}
   and 
   \begin{eqnarray}                          
   V_2 = V_2(x_1,x_2)  
       \nonumber \\          
   = [m]_4   + 4 [m]_3  (k_1 x_1 + k_2 x_2)                   
       + 6 [m]_2 \left( [k_1]_2 x^2_1  + 2 k_1 k_2 x_1 x_2 + [k_2]_2  x^2_2 \right)
                                                               \nonumber \\
      + 4 m \left( [k_1]_3 x^3_1  +  3 [k_1]_2 k_2  x^2_1 x_2 
           +  3 k_1 [k_2]_2  x_1 x_2^2 +  [k_2]_3 x^3_2 \right)  
                                                                \nonumber \\   
       + [k_1]_4 x^4_1 +  4 [k_1]_3 k_2  x^3_1 x_2                                                                 
             + 6 [k_1]_2 [k_2]_2  x^2_1 x_2^2
             + 4 k_1 [k_2]_3  x_1 x_2^3 + [k_2]_4 x^4_2.       \label{3.12b}                              
                                                         \end{eqnarray} 
Here  $[N]_k = N (N-1)... (N - k +1)$.

Due to (\ref{3.9}) we obtain 
\begin{equation}
  \lambda = \alpha \Lambda 
     = - \frac{V_1}{4 {\cal P} }  +  \frac{V_2}{8 {\cal P}^2 }, 
                         \label{3.13}    
   \end{equation} 
or 
\begin{equation}
  V_2(x_1,x_2) - 2 {\cal P}(x_1,x_2) V_1 (x_1,x_2)  -   8 ({\cal P}(x_1,x_2))^2 \lambda = 0. 
                         \label{3.14}    
 \end{equation} 
 Owing to eq. (\ref{3.15}) we get
\begin{equation}
 x_2 =  x_2(x_1) = - \frac{m -1}{k_2 -1}  -  \frac{k_1 - 1}{k_2 -1} x_1  \label{3.16}
\end{equation}
 Hence, from eq. (\ref{3.14})  we  get a master equation in $x_1$ variable
 \begin{equation}
   V_2(x_1,x_2(x_1)) - 2 {\cal P}(x_1,x_2(x_1)) V_1 (x_1,x_2(x_1)) 
    -   8 ({\cal P}(x_1,x_2(x_1)))^2 \lambda = 0. 
                          \label{3.17}    
  \end{equation}  
This polynomial equation is of fourth order or less (this depends upon the value of $\lambda$). 
One may solve it in radicals for all $m > 1$, $k_1 > 1$ and $k_2 > 1$. 

Relations  (\ref{3.10b}) and (\ref{3.16}) imply the identity
 \begin{eqnarray}
 - (k_2 - 1) {\cal P}(x_1,x_2(x_1)) = (k_1 - 1)(k_1 + k_2 -2)x_1^2  \nonumber \\ 
   + 2(m-1)(k_1 -1)x_1 + (m-1)(m + k_2 - 2),
 \label{3.18} 
\end{eqnarray}
which will be used below.

\section{The case $k_1  \neq k_2$}

In this section  we put  $k_1  \neq k_2$.  
We rewrite  relation (\ref{3.13}) as following
\begin{equation}
  \lambda = f(x_1) \equiv   
      - \frac{V_1 (x_1,x_2(x_1))}{4 {\cal P}(x_1,x_2(x_1))} 
      +  \frac{V_2(x_1,x_2(x_1))}{8 ({\cal P}(x_1,x_2(x_1)))^2 }. 
                         \label{3.13a}    
   \end{equation} 

Due to (\ref{3.16}) we present  restrictions (\ref{3.3a}) in the following form \cite{ErIv-20}
 \begin{equation}
        x_1 \neq X_1, \quad  x_1 \neq X_2, \quad  x_1 \neq X_3, \quad  x_1 \neq X_4,
      \label{3.3b}
    \end{equation}
where
 \begin{eqnarray}
 X_1 = 1,
   \label{3.x1} \\
 X_2 = -\frac{m + k_2 -2}{k_1 -1},
   \label{3.x2} \\
 X_3 = -\frac{m-1}{k_1 + k_2 -2},
   \label{3.x3} \\
 X_4 = \frac{m - k_2}{k_2 - k_1}.
   \label{3.x4}
 \end{eqnarray}

\subsection{ Extremum points }

We obtain \cite{ErIv-20}  

\begin{equation}
\frac{df}{dx_1}= \frac{C(m, k_1, k_2)(x_1 - X_1)(x_1 - X_2)(x_1 - X_3)(x_1 - X_4)}{\bigg(
  - (k_2 - 1) {\cal P}(x_1,x_2(x_1)) \bigg)^3},
  \label{3.f}
 \end{equation}
 where
 \begin{equation}
 C(m, k_1, k_2) = (m-1)(k_1 - 1)^2(k_2 - k_1)(k_1 + k_2 -2)
  \label{3.c}
 \end{equation}
  and $X_1, X_2,  X_3, X_4$ are given by (\ref{3.x1})-(\ref{3.x4}). Thus, the 
  extreme points of the function  $f(x_1)$ are excluded from our consideration 
  Due to   (\ref{3.3}) we are ought to exclude the  extreme points of   $f(x_1)$.
 
For  $\lambda_i = f(X_i)$, $i =1,2,3,4$, we have \cite{ErIv-20}

\begin{eqnarray}
\lambda_1 = \lambda_1 (m,k_1,k_2) = \frac{u(k_2,m + k_1)}{8(m + k_1 + k_2 - 3)(m + k_1 -2)(k_2 -1)},
   \label{3.l1L} \\
\lambda_2 = \lambda_2 (m,k_1,k_2) = \frac{ u(k_1, m + k_2)}{8(m + k_1 + k_2 - 3)(m + k_2 -2)(k_1 -1)},
   \label{3.l2L} \\
\lambda_3 = \lambda_3 (m,k_1,k_2) = \frac{u(m, k_1 + k_2)}
{8(m-1) (k_1 + k_2 -2)(m + k_1 + k_2 - 3)},
   \label{3.13L} \\
 \lambda_4 = \lambda_4 (m,k_1,k_2) = \frac{v(m,k_1,k_2)}{8 w(m,k_1,k_2)}.
   \label{3.l4L}
 \end{eqnarray}
Here
\begin{eqnarray}
u(m, l) = l m^2 + (l^2 - 8l +8)m + l(l -1), \label{3.u} \\
v(m,l,k) = (k+ l)m^2 + (m + l )k^2 +(m + k)l^2 - 6mlk, \label{3.v} \\
w(m,l,k) = (k+ l - 2)m^2 + (m + l - 2)k^2 +(m + k - 2)l^2
\nonumber \\
   + 2 m l + 2mk + 2lk   -  6mlk. \label{3.w}
\end{eqnarray} 

It was verified in Ref. \cite{ErIv-20} that
 \begin{equation}
   \lambda_i = \lambda_i(m,k_1,k_2)> 0 \label{3.L} 
 \end{equation}
for $m > 1$, $k_1 > 1$, $k_2 > 1$, $i = 1,2,3,4$. 

In the limit $x_1 \to \pm \infty $ we obtain
\begin{equation}
  \lambda_{\infty}= \lim_{x_1 \to  \infty } f(x_1) =   
  \frac{(k_1 + k_2 -6)k_1k_2 + k_1^2 + k_2^2 + k_1 + k_2}{8(k_1-1)(k_2-1)(k_1 + k_2 -2)}.
  \label{3.in}
\end{equation}

 Here we obtain \cite{ErIv-20} 
\begin{equation}
  \lambda_{\infty} = \lambda_{\infty}(k_1, k_2) = \lambda_{\infty}(k_2, k_1) > 0,
  \label{3.inn}
\end{equation}
for all $k_1 > 1$ and $k_2 > 1$.

The definitions of $X_i$ imply  \cite{ErIv-20}
 \begin{equation}
       X_2 <  X_3 < 0 < X_1 = 1.
   \label{3.X123}
 \end{equation}
Here $m > 1$, $k_1 > 1$ and $k_2 > 1$.

From this point up to Section 4 we impose the following 
inequality
 \begin{equation}
   1 < m < k_1 < k_2.
  \label{3.mk1k2}
 \end{equation}

It was shown in Ref. \cite{ErIv-20} that
\begin{equation}
   0 < \lambda_1 < \lambda_2 < \lambda_3,
  \label{3.Lmk1k2}
 \end{equation}
\begin{equation}
   0 < \lambda_1 < \lambda_4 < \lambda_3.
  \label{3.Lm4k2}
 \end{equation}
and
 \begin{equation}
  (A_{+}) \quad X_4 < X_2, \qquad  \lambda_4 > \lambda_2, 
  \quad {\rm for} \quad 2k_1 - m - k_2 > 0,
  \label{3.XL1}
 \end{equation}

\begin{equation}
  (A_{-}) \quad X_4 > X_2, \qquad  \lambda_4 < \lambda_2, 
  \quad {\rm for} \quad 2k_1 - m - k_2 < 0,
  \label{3.XL2}
 \end{equation}
and 
\begin{equation}
  (A_{0}) \quad X_4 = X_2, \qquad  \lambda_4 = \lambda_2, 
  \quad {\rm for} \quad 2k_1 - m - k_2 = 0.
  \label{3.XL3}
 \end{equation}

For $(m, k_1, k_2) = (4, 6, 7)$ the  the function $\lambda = f(x_1)$ is presented grafically 
 at Figure 1.

 \begin{figure}[!h]
	\begin{center}
		\includegraphics[width=0.75\linewidth]{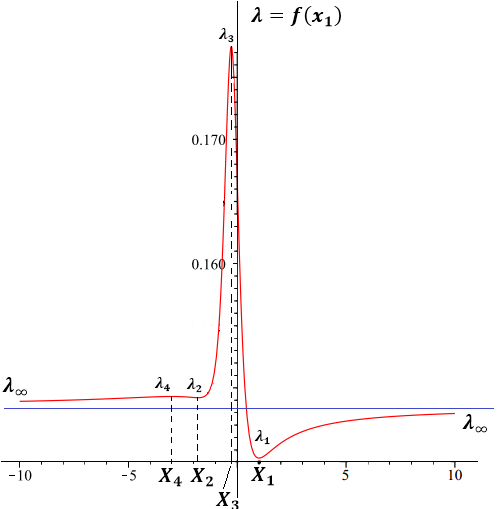}
		\caption{The grafical representation of the
		function $\lambda = f(x_1)$ for  $m=4$,  $k_1=6$, $k_2 = 7$ \cite{ErIv-20}.}
		\label{rfig:1}
	\end{center}
 \end{figure}

It was proved in Ref. \cite{ErIv-20}   that
\begin{equation}
 \lambda_1 < \lambda_{\infty} < \lambda_3.
 \label{3.L1inf3}
\end{equation}

By using  (\ref{3.mk1k2}) and (\ref{3.c}) we get
\begin{equation}
 C(m, k_1, k_2) >0. 
  \label{3.cc}
 \end{equation}
 

  It was proved in Ref. \cite{ErIv-20}
  that for the function $f(x_1)$ mentioned above  
 $X_3$ is the point of absolute maximum and $X_1$ is the point of absolute minimum,
 i.e.    
 \begin{equation}
       \lambda_1   \leq   \lambda = f(x_1) \leq \lambda_3
   \label{3.Lminmax}
  \end{equation}
 for all $x_1 \in \R$. 
 We remind that according to (\ref{3.3b}) the points  $X_1, X_2, X_3, X_4$ are forbidden for our
 analysis. We obtain 
 \begin{equation}
 \lambda_1   < \lambda = f(x_1) < \lambda_3 
  \label{3.Lminmaxno}
 \end{equation}
 for all $x_1 \neq X_1, X_2, X_3, X_4$. 
 Let us denote the set of definition of the fuction $f$
 for our consideration
 $(-\infty, \infty)_{*} \equiv \{x| x \in \R, x \neq X_1, X_2, X_3, X_4 \}$.
 Since the function $f(x_1)$ 
 is continuous one the image of the function $f$ 
 (due to intermediate value theorem) is 
 $f((-\infty, \infty)_{*}) = (\lambda_1,  \lambda_3)$. 
 Thus, we a led the following proposition.
 
 {\bf Proposition 1.} {\em  The solutions to equations
 (\ref{2.3}), (\ref{2.4})  for  ansatz  (\ref{3.1}) with   $1 < m < k_1 < k_2$ 
 obeying the inequalities $H \neq 0$,  $H \neq h_1$, $H \neq h_2$, $h_1 \neq h_2$ and 
 $S_1 = m H + k_1 h_1 + k_2 h_2 \neq 0$ do exist if and only if $\alpha \eps> 0$ and 
 \begin{equation}
    0 < \lambda_1 <  \alpha \Lambda < \lambda_3,
   \label{3.L13}
  \end{equation}
   where $\lambda_1$ and $\lambda_3$ 
  are defined in (\ref{3.l1L}) and (\ref{3.13L}), respectively. 
  In this case  $x_1 = h_1/H \neq X_1, X_2, X_3, X_4$ 
 (see (\ref{3.x1}), (\ref{3.x2}), (\ref{3.x3}), (\ref{3.x4})),  
  $x_2 = h_2/H = x_2(x_1)$ is given by  (\ref{3.16}), $x_1$ obeys the 
  polynomial master equation (\ref{3.17}) (of fourth order or less) and 
  $H^2$ is given by (\ref{3.9}) and (\ref{3.10}). } 
 
 {\bf The case $H = 0$.} 
 In the case $H = 0$ the solutions under consideration  
 take place only if $\alpha \eps > 0$, $\Lambda \eps > 0$  and 
   \begin{equation}
    \alpha \Lambda = \lambda_{\infty}(k_1,k_2) 
    = \frac{(k_1 + k_2 -6)k_1k_2 + k_1^2 + k_2^2 + k_1 + k_2}{8(k_1-1)(k_2-1)(k_1 + k_2 -2)} > 0, 
   \label{3.R1.0}
   \end{equation}
 where $k_1 \neq k_2$. Indeed, relation (\ref{3.4L}) reads as  
 $(k_1 - 1) h_1 + (k_2 - 1) h_2 = 0$,                                        
 and relation (\ref{3.4Q}) is equivalent to   
  $(k_1 - 1) (h_1)^2 + (k_2 - 1) (h_2)^2 = 1/(2 \alpha \eps)$.
 From these relations we get  $\alpha \eps > 0$ and
 \begin{eqnarray}
 h_1 = \pm \left(\frac{k_2 -1}{2 \alpha \eps (k_1 -1) (k_1 + k_2 - 2)}\right)^{1/2},
    \label{3.R1.1} \\ 
 h_2 = \mp \left(\frac{k_1 -1}{2 \alpha \eps (k_2 -1) (k_1 + k_2 - 2)}\right)^{1/2}
   \label{3.R1.2},
 \end{eqnarray}
 which imply, due to $H=0$ and (\ref{3.4E}),
 the relation  (\ref{3.R1.0}).  
 
\section{The case $k_1 = k_2$}

We will now turn our attention to the case $H \neq 0$, $m > 1$ and $k_1 = k_2 = k > 1$.
Due to (\ref{3.15}) we obtain 
\begin{equation}
    m -1  + (k - 1)(x_1 +  x_2) = 0.  \label{4.1}
\end{equation}
It follows from (\ref{3.10b}) that
\begin{equation}
 {\cal P}  = 1 - m + (1 - k) (x_1^2 +  x_2^2). 
 \label{4.2}
\end{equation} 

Since the case of equal factor-space dimensions 
is excluded from our consideration (see Section 2)
we put 
\begin{equation}
    m \neq k  \label{4.1m}
\end{equation}
and $\alpha \eps > 0$.

Denoting 
\begin{equation}
   X \equiv \alpha \eps H^2,    \label{4.3}
\end{equation}
$\alpha \eps > 0$, we obtain from
(\ref{3.9}) that
\begin{equation}
   X  {\cal P} = - \frac{1}{2}.    \label{4.4}
\end{equation}
Relation (\ref{4.3}) implies 
  \begin{equation}
     H = \epsilon_0 \sqrt{X/\alpha \eps}, \qquad \epsilon_0 = \pm 1 . 
        \label{4.5}
  \end{equation}

Plugging the relations (\ref{4.1}), (\ref{4.2})
into (\ref{3.12a}), (\ref{3.12b}) we obtain
\begin{eqnarray}
V_1 = [(m-1)(m-k) + {\cal P} k (k-1)]/(k-1)^2,     \label{4.6} \\  
V_2  = [- (m-1)(m-k)(m+k-2)(m+2k-3)  \nonumber \\
         + 3 {\cal P}^2 (k-1)^2 k  ]/(k - 1)^3.       \label{4.7}
\end{eqnarray}
  By virtue  relation (\ref{4.4}) we present relation (\ref{3.13}) as
  \begin{equation}
  2 \lambda = 2 \alpha \Lambda  =  X V_1 +  X^2 V_2, 
                              \label{4.8}    
  \end{equation} 
 or in equivalent manner as  
 \begin{equation}
  A X^2 + B X + C = 0.          \label{4.9}
  \end{equation} 
  Here
  \begin{eqnarray}
 A = (m-1)(m-k)(m+k-2)(m+2k-3), \label{4.9A} \\
 B = -(m-1)(m-k)(k-1),   \qquad      \label{4.9B} \\
 C=  - \frac{1}{4} k(k-1)^2 + 2 \lambda (k - 1)^3. \quad  \label{4.9C} \\  
 \end{eqnarray}
It follows from (\ref{4.1m}) that $A \neq 0$. 
The calculation of the discriminant $D = B^2 - 4 A C$ leads us to the 
following identity
 \begin{equation}
   D =  (m-1)(m-k)(k-1)^2 (F   - 8 \lambda f),   \label{4.10}
   \end{equation} 
  where we denote
 \begin{eqnarray}
   F = F(m,k) = (m-1)(m-k) + (m+ k-2)(m+2k-3)k, \label{4.11}      \\
   f = f(m,k) = (m+k-2)(m+2k-3)(k - 1) > 0  \label{4.12} . 
  \end{eqnarray}
It was veridied in Ref. \cite{ErIv-20} that
$F = F(m,k) >0$ for all $m > 1$, $k > 1$ and $k \neq m$. 

By solving  eq. (\ref{4.9}) we get \cite{ErIv-20}
\begin{equation}
  X = (- B + \bar{\epsilon}_1 \sqrt{D})/(2A), \qquad \bar{\epsilon}_1 = \pm 1. 
        \label{4.14}
 \end{equation}

We seek real solutions obeying   
\begin{eqnarray}
 D  > 0, \label{4.15D} \\
 X > 0.   \label{4.15X} 
 \end{eqnarray}
   The case $D = 0$ should be excluded \cite{ErIv-20}. Indeed, $D = 0$ 
  implies either $x_1 = 1$ or $x_2 = 1$, which is in contradiction with (\ref{3.3a}).

 Here we rewrite the inequality (\ref{4.15D}) as 
 \begin{eqnarray}
  \lambda  < \lambda_1 \ {\rm for} \ m > k,   \label{4.16a} \\
  \lambda  > \lambda_1 \ {\rm for} \ m < k,   \label{4.16b}   
  \end{eqnarray}
 where 
 \begin{equation}
   \lambda_1 = \lambda_1(m,k,k) = F(m,k)/(8 f(m,k)). 
         \label{4.17}
  \end{equation}

Equations (\ref{4.1}) and (\ref{4.2}) 
may be resolved as 
\begin{eqnarray}
x_1 = -(\epsilon_2 \sqrt{E}   
      +m-1)/(2k-2), \label{4.18a} \\
x_2 = -(- \epsilon_2 \sqrt{E} + m-1)/(2k-2), 
\label{4.18b}
\end{eqnarray}
where $\varepsilon_2 = \pm 1$ and 
\begin{eqnarray}
  E = -(m-1)(m+2k-3) - 2 {\cal P} (k - 1)  \nonumber \\
    =  (k-1) X^{-1} - (m-1)(m+2k - 3). 
  \label{4.19}
\end{eqnarray}
Here one should consider the case  
 \begin{equation}
   E > 0  \label{4.20}.
  \end{equation}
Indeed, $E = 0$ implies  $x_1 = x_2$ 
which is not allowed by  (\ref{3.3a}). Due to 
(\ref{4.15X}) and (\ref{4.20}) we obtain 
\begin{equation}
   0 < X < \frac{k-1}{(m-1)(m+2k - 3)}.  \label{4.21}
  \end{equation}

  
  It was verified in Ref. \cite{ErIv-20}  that relations 
  (\ref{4.18a}), (\ref{4.18b}) and  (\ref{4.21}) imply
  all four inequalities in (\ref{3.3a}).
     
   Now we proceed with inequalities in (\ref{4.21}). By introducing 
   the parameter 
   \begin{equation}
          \epsilon_1 = \bar{\epsilon}_1 {\rm sign} (m-k),  \label{4.26}
   \end{equation}
   we rewrite relation (\ref{4.14}) in the following form
 \begin{equation}
     X  = \frac{k-1}{2 (m + k -2)(m+2k - 3)} 
           + \epsilon_1 \frac{\sqrt{D}}{2|A|},         \label{4.27}
 \end{equation}
 $\epsilon_1 = \pm 1$. 

First, we consider the case  $\epsilon_1 = - 1$. 
The second inequality in  (\ref{4.21})
$X < \frac{k-1}{(m-1)(m+2k - 3)}$ is valid due to $2 (m + k -2) > m - 1$. 
 As to the first inequality $X > 0$, we obtin 
 \begin{equation}
     0 < \sqrt{D} < (m-1)|m-k|(k -1).         \label{4.28}
  \end{equation}
  Due to definition of $D$ in (\ref{4.10}) we get 
  \begin{equation}
    0 < (m-1)(m-k)(k -1)^2 (F - 8 \lambda f) <  (m-1)^2 |m-k|^2 (k - 1)^2.  \label{4.29}
  \end{equation} 
 
 Relations  (\ref{4.29}) may be presented in following form 
 \begin{eqnarray}
   F_{-} < 8 \lambda f <  F,  \ {\rm for} \  m > k,   \label{4.30a} \\
   F < 8 \lambda f <  F_{-}, \ {\rm for} \  m < k.   \label{4.30b} 
 \end{eqnarray} 
 Here
 \begin{equation}
     F_{-} \equiv   F - (m-1)(m-k).  \label{4.30c}
  \end{equation}
 
 By using relations 
   \begin{equation}
      \frac{F_{-}}{8f} = \frac{k}{8(k-1)} = \lambda_{\infty} = \lambda_{\infty}(k,k),
                         \label{4.30de}
   \end{equation}
 where $\lambda_{\infty}(k,l)$ is defined in  (\ref{3.in}), and
 (\ref{4.17}) and (\ref{4.30de}) one can present relations (\ref{4.30a}), (\ref{4.30b})
 in the following form
 \begin{eqnarray}
    \lambda_{\infty} <  \lambda  <   \lambda_{1},  \ {\rm for} \  m > k,   \label{4.30aa} \\
    \lambda_{1} <  \lambda  <  \lambda_{\infty}, \ {\rm for} \  m < k.   \label{4.30bb} 
  \end{eqnarray}

 Now, we consider the case  $\epsilon_1 = 1$. Since the inequality $X > 0$ is obeyed 
 this case, one should verify the inequality $X < \frac{k-1}{(m-1)(m+2k - 3)}$.
 We find 
 \begin{equation}
   0 < \sqrt{D} < |m-k|(m+2k-3)(k-1),     \label{4.31}
 \end{equation}
 or
 \begin{equation}
    0 < (m-1)(m-k) (F   - 8 \lambda f) <  |m-k|^2 (m+2k-3)^2.  
              \label{4.32}
   \end{equation}
 We write relations  (\ref{4.32}) in the following form 
  \begin{eqnarray}
   F_{+}  < 8 \lambda f <  F,  \ {\rm for} \  m > k,   \label{4.33a} \\
   F < 8 \lambda f <  F_{+} , \ {\rm for} \  m < k,   \label{4.33b} 
  \end{eqnarray}
where 
\begin{equation}
  F_{+} \equiv   F - (m-1)^{-1}(m-k)(m+2k-3)^2.          \label{4.33c}
 \end{equation}

Here one can  verify  that 
   \begin{equation}
      \frac{F_{+}}{8f} = \lambda_{3} = \lambda_{3}(m,k,k).
                         \label{4.33de}
   \end{equation}
 Due to (\ref{4.17}) and (\ref{4.33de}) we rewrite relations 
 (\ref{4.33a}), (\ref{4.33b})  in the following form
 \begin{eqnarray}
    \lambda_{3}  <  \lambda  <  \lambda_{1},  \ {\rm for} \  m > k,   \label{4.33aa} \\
    \lambda_{1} <  \lambda  <  \lambda_{3} , \ {\rm for} \  m < k .  \label{4.33bb} 
   \end{eqnarray}
 
 Here   
 \begin{equation}
  \lambda_{1} <  \lambda_{\infty}  <  \lambda_{3}
                          \label{4.33dd}
  \end{equation}
 for $m < k$, while 
 \begin{equation}
   \lambda_{3} <  \lambda_{\infty}  <  \lambda_{1}
                           \label{4.33ee}
 \end{equation}
 for $k < m$. The inequalities in (\ref{4.33ee}) 
 just follow from inequalities $F_{+} < F_{-} < F$ for $k < m$. 
 
 Thus, we are led to the following  generalisation of 
 the Proposition 2 from Ref. \cite{ErIv-20}.
  {\bf Proposition 2.} {\em  The solutions to Eqs.
  (\ref{2.3}), (\ref{2.4})  for  ansatz  (\ref{3.1}) imposed with   $1 < m$, $1 < k_1 = k_2 = k$,
  $m \neq k$, 
  obeying the inequalities $H \neq 0$,  $H \neq h_1$, $H \neq h_2$, $h_1 \neq h_2$, 
  $S_1 = m H + k h_1 + k h_2 \neq 0$ do exist if and only if $\alpha \eps > 0$, 
  \begin{equation}
     \lambda_1 < \lambda = \alpha \Lambda < \lambda_3
    \label{4.L13}
   \end{equation}
   for $m < k$  and 
   \begin{equation}
     \lambda_3 < \lambda = \alpha \Lambda < \lambda_1,
         \label{4.L31}  
    \end{equation}
   where $\lambda_1 = \lambda_1(k,k)$,  $\lambda_3=\lambda_3(k,k)$ 
   are defined in (\ref{3.l1L}) and  (\ref{3.13L}). 
   In this case $H$ satisfies the relation (\ref{4.5}) with $X$ from 
      (\ref{4.27}),  $x_1 = h_1/H$ and 
   $x_2 = h_2/H$ are given by relations (\ref{4.18a}) and (\ref{4.18b}),
   $\lambda$ obeys  (\ref{4.30aa}),  (\ref{4.30bb})   for  $\epsilon_1 = - 1$
   and (\ref{4.33aa}), (\ref{4.33bb}) for  $\epsilon_1 =  1$
     with  $\lambda_{\infty} = \frac{k}{8(k-1)}$.  }

 {\bf The case $H = 0$.} For $k_1 = k_2 = k > 1$ and  $H = 0$ the solutions under consideration
  obeying restrictions (\ref{3.3}) are absent \cite{ErIv-20}.

 \section{The analysis of stability}

Here we analyse the stability of our solutions along a line as it was 
done in refs. \cite{ErIvKob-16,Ivas-16,ErIv-17-2}.

We impose  the following restriction 
\begin{equation}
  \det (L_{ij}(v)) \neq 0,
  \label{5.2}
\end{equation}
where
\begin{equation}
L =(L_{ij}(v)) = (2 G_{ij} - 4 \alpha \eps G_{ijks} v^k v^s).
   \label{5.1b}
 \end{equation}

Here one should deal with general cosmological type setup with the metric 
\begin{equation}
 g= - \eps du \otimes du + \sum_{i=1}^{n} e^{2\beta^i(u)} \eps_i dy^i \otimes dy^i,
 \label{5.3}
\end{equation}
where $\eps = \pm 1$, $\eps_i = \pm 1$, $i = 1, \dots, n$. 
For the  equations of motion we obtain  \cite{IvKob-18mm} 
\begin{eqnarray}
     E = G_{ij} h^i h^j + 2 \Lambda \eps - \alpha \eps G_{ijkl} h^i h^j h^k h^l = 0,
         \label{5.3.1} \\
         Y_i =  \frac{d L_i}{dt}  +  (\sum_{j=1}^n h^j) L_i -
                 \frac{2}{3} (G_{sj} h^s h^j -  4 \Lambda \eps) = 0,
                     \label{5.3.2a}
          \end{eqnarray}
where $h^i = \dot{\beta}^i = \frac{d \beta^i}{du}$,           
 \begin{equation}
  L_i = L_i(h) = 2  G_{ij} h^j
       - \frac{4}{3} \alpha \eps G_{ijkl}  h^j h^k h^l  
       \label{5.3.3},
 \end{equation}
 $i = 1,\ldots, n$.

According to previous consideration of Ref. \cite{Ivas-16}  the solution
$(h^i(t)) = (v^i)$ ($i = 1, \dots, n$; $n >3$) to eqs. (\ref{5.3.1}), (\ref{5.3.2a})
which obeys the restrictions  (\ref{5.2}) is  stable under perturbations
\begin{equation}
 h^i(t) = v^i +  \delta h^i(t), 
\label{5.3h}
\end{equation}
 $i = 1,\ldots, n$,  as $u \to + \infty$, if and only if
 \begin{equation}
   S_1(v) = \sum_{i = 1}^{n} v^i >0
   \label{5.1s}
 \end{equation}
 and it is unstable, as $u \to + \infty$, if and only if
 \begin{equation}
    S_1(v) = \sum_{i = 1}^{n} v^i < 0.
    \label{5.1ns}
  \end{equation}

In the limit  $u \to - \infty$ the stability condition is given by (\ref{5.1ns}) 
while the unstability condition reads as (\ref{5.1s}). These conditions 
just follow from  solutions for perturbations
$\delta h^i(t) = C_i \exp(- S_1(v) u)$ ($C_i = {\rm const} \neq 0$)
which are valid in the leading order.

 Here a key point is the verification of the relation (\ref{5.2}). It was fulfilled 
 in Ref.  \cite{ErIv-17-2} by using first three relations in (\ref{3.3}) 
 and (\ref{3.5b}) and $k_1 > 1$, $k_2 > 1$ and $m >1$.
 
     First we consider the case  $1 < m < k_1 < k_2$. By using (\ref{3.15})
   we find that for $H > 0$ the  condition (\ref{5.1s}) 
   may be written as
    \begin{equation}
       m  + k_1 x_1 + k_2 x_2 = 1  +  x_1 + x_2 >0
      \label{5.1xs}
    \end{equation} 
    or, equivalently, 
    \begin{equation}
            x_1 > X_4 = \frac{m - k_2}{k_2 - k_1}.
          \label{5.1xxs}
     \end{equation}
   For $H < 0$    the stability  condition  (\ref{5.1s})  
   is following one  
         \begin{equation}
                 x_1 < X_4.
               \label{5.1xxsa}
          \end{equation}   

The non-stability condition (\ref{5.1ns}) for $u \to + \infty$ reads as 
(\ref{5.1xxsa}) for $H > 0$ and as (\ref{5.1xxs})
for $H < 0$.  These conditions 
 are reversed in case  $u \to - \infty$.

{\bf Proposition 3.} {\em  Let us consider cosmological type solutions
 to equations   (\ref{2.3}), (\ref{2.4})  for  ansatz  (\ref{3.1}) with   $1 <  k_1 < k_2$,
  obeying the inequalities $H \neq 0$,  $H \neq h_1$, $H \neq h_2$, $h_1 \neq h_2$, 
  $S_1 = m H + k_1 h_1 + k_2 h_2 \neq 0$. \\
   (a) Let $H > 0$. For $u \to + \infty$ the solutions are  stable  
       if  $x > X_4$ and unstable if  $x < X_4$, while
       for $u \to - \infty$ they are stable if  $x < X_4$ and unstable if  $x > X_4$. \\
   (b) Let $H < 0$. For $u \to + \infty$ the solutions are  stable  
         if  $x < X_4$ and unstable if  $x > X_4$, while
         for $u \to - \infty$ they are stable if  $x > X_4$ and unstable if  $x < X_4$.  
    } 

Now we proceed with considering the case $H \neq 0$, $1 < m$, $1 < k_1 = k_2 = k$,  $m \neq k$. 
Since $x_1 \neq 1$, $x_2 \neq 1$ and $x_1 \neq x_2$ the exact 
solutions under consideration  obey first three relations in (\ref{3.3}) 
which imply   the validity of the key restriction (\ref{5.2}).

 For the stability condition (\ref{5.1s}) as $u \to + \infty$ in this case we get, 
    \begin{equation}
        H ( m  + k_1 x_1 + k_2 x_2) = H (1  +  x_1 + x_2) = 
        H \left(1 - \frac{m - 1}{k - 1}\right) > 0,
        \label{5.2xs}
    \end{equation} 
 or, equivalently,  
 \begin{equation}
        H (k - m) > 0.
        \label{5.2xss}
    \end{equation}

 The non-stability condition (\ref{5.1ns}) for $u \to + \infty$ 
 may be written as
   \begin{equation}
         H (k - m) < 0.
         \label{5.2nxs}
     \end{equation} 

Thus, we have  the following proposition.

{\bf Proposition 4.} {\em  Let us consider cosmological type solutions
  (\ref{2.3}), (\ref{2.4})  for  ansatz  (\ref{3.1}) with   $1 < m$, $1 < k_1 = k_2 = k$,
  $m \neq k$,   obeying the inequalities $H \neq 0$,  $H \neq h_1$, $H \neq h_2$, $h_1 \neq h_2$, 
  $S_1 = m H + k h_1 + k h_2 \neq 0$,
   is stable, as $u \to + \infty$, if and only if 
   $ H (k - m) > 0$  and it is unstable, as $u \to + \infty$, 
      if and only if  $H (k - m) < 0$. \\
      (c) Let $H > 0$. For $u \to + \infty$ the solutions are  stable  
             if  $k  > m$ and unstable if  $k < m$, while
             for $u \to - \infty$ they are stable if  $k < m$ and unstable if  $k > m$. \\
         (d) Let $H < 0$. For $u \to + \infty$ the solutions are  stable  
               if  $k < m$ and unstable if  $k > m$, while
               for $u \to - \infty$ they are stable if  $k > m$ and unstable if  $k < m$
       } 

{\bf The case $H = 0$.} For a completeness we consider the solutions with $H=0$ and  $h_1$, $h_2$ 
from (\ref{3.R1.1}), (\ref{3.R1.2}), where $k_1 \neq k_2$, $k_1 > 1$,  $k_2 > 1$, 
  $\alpha \eps > 0$ and  $\Lambda$ is given by (\ref{3.R1.0}).   We get 
  \begin{equation}
    S_1 = k_1 h_1 + k_2 h_2 =  \pm (k_2 - k_1)
    \left(2 \alpha \eps (k_1 -1)(k_2 -1) (k_1 + k_2 - 2) \right)^{-1/2}.
    \label{5.S1}  
 \end{equation}
 Here $\pm$ is a sign parameter in (\ref{3.R1.1}), (\ref{3.R1.2}).
 By using our analysis presented above we obtain that the solution with  
 $\pm (k_2 - k_1) > 0$ is stable, as $u \to + \infty$. 
 This occurs if either  $k_2 > k_1$ and the sign $``+"$ is selected in (\ref{3.R1.1}) and (\ref{3.R1.2}), 
 or if $k_2 < k_1$ and the sign $``-"$ is chosen. 
  For the case $\pm (k_2 - k_1) < 0$ our solution is unstable, 
  as $u \to + \infty$.  (Here we also assume the restriction $m > 1$.)  These conditions 
   are reversed in case  $u \to - \infty$.

\section{Conclusions}

We have studied the  $D$-dimensional  Einstein-Gauss-Bonnet (EGB) model with the $\Lambda$-term and two non-zero constants $\alpha_1$ and $\alpha_2$.   By dealing with diagonal  cosmological type  metrics, 
we have considered a class of solutions with  exponential dependence of three scale factors (upon $u$-variable)  for any 
$\alpha = \alpha_2 / \alpha_1  \neq 0$, signature parameter $\eps = \pm 1$ and   generic dimensionless parameter 
$\Lambda \alpha$.

More precisely  speaking  we have described   a class of  cosmological type  solutions with  exponential dependence of three scale factors, governed by three non-coinciding  Hubble-like parameters $H$, $h_1$ and $h_2$. These parameters correspond, respectively,  to factor spaces  of dimensions $m > 1$,  $k_1 > 1$   and $k_2 > 1$ 
($D = 1 + m + k_1 + k_2$),  and obey the following restriction  $S_1 = m H + k_1 h_1 + k_2 h_2 \neq 0$.  
We have analyzed two cases: i) $ m < k_1 < k_2$ and ii)  $1< k_1 = k_2 = k \neq m$. This choice does not restrict the generality, since, as it was shown,  there are no solutions under consideration for $k_1 = k_2 = m$.) It was shown that  the solutions  exist only if,  $\lambda = \alpha \Lambda > 0$ and  the  (dimensionless) parameter  $\lambda$ obeys certain restrictions, e.g. upper and lower bounds for $H \neq 0$,  which depend upon  dimensions $m$, $k_1$ and $k_2$  (Proposition 1).   In the case ii) we have presented  explicit  solutions for all $k > 1$ and $k \neq m$  ( Proposition 2). 

By using   Chirkov-Pavluchenko-Toporensky splitting trick from Ref. \cite{ChPavTop1}  we have  reduced  the  problem for $H \neq 0$ to  master equation    on the dimensionless variable  $x_1  = h_1/H$. This  equation is of fourth order (in generic case) or less (depending on  $\lambda$), and  may be solved in radicals for all  $m > 1$, $k_1 > 1$, $k_2 > 1$ and $\lambda $. The master equation does not depend upon the signature parameter $\eps = \pm 1$ which is only controlling the sign  of $\alpha $ according to  inequality $\alpha \eps  > 0$. Due to bounds obtained $\lambda = \alpha \Lambda >0 $. (This is valid also for $H=0$). Hence the solutions under consideration do exist if $ \Lambda  \eps  >0$ , i.e. when   
$ \Lambda   >0$ in cosmological case ($ \eps  = 1$) and $ \Lambda   < 0$ in static case ($ \eps  = - 1$). 
Here there are no solutions under considerations  for  $ \Lambda = 0 $ - contrary to the case of two
factor spaces \cite{I-20,IvKob-19-EPJC}.  

Here we have analyzed  the stability of  solutions as  $u \to \pm \infty$ in a class of cosmological type solutions with diagonal  metrics. In both cases ((i) and (ii)) for $H \neq 0$ the ``islands'' of stability and instability   were singled out. (The case $H=0$ was also analysed.) We have shown that  in the case i) the solutions with $H > 0$ are stable as $u \to  \infty$ for $x_1 = h_1/H  > X_4 = \frac{m - k_2}{k_2 - k_1}$ and  unstable as $u \to  \infty$ for $x_1 < X_4$ (see Proposition 3). 
These conditions  should be reversed when we consider the case $H > 0$,  $u \to - \infty$ or  we deal with $H < 0$,  $u \to + \infty$ (see Proposition 3).  It was proved  that in the case ii) the solutions with $H > 0$  are stable as  $u \to  \infty$ for $k > m$ and unstable as  $u \to  \infty$ for $k < m$ (see Proposition 4).  For given choice of asymptotic $u \to \pm \infty$ the stability condition for $H < 0$ is equivalent to instability conditions for $H >0$ and vice versa. 

We have also found  that the solution with $H =0$  exists only for $k_1 \neq k_2$, $\alpha \eps > 0$ and fixed value of $\eps \Lambda > 0$ depending upon $k_1$ and $k_2$.  Here we have two   opposite in sign  solutions  for $(h_1, h_2)$ with  one solution being stable ($u \to \pm \infty$) and the second one - unstable depending upon the sign of  $k_1 - k_2$.

Some cosmological applications of the model ($\eps = 1$), e.g. in context of variation of gravitational constant, where considered in Refs. \cite{ErIv-19-GC,ErIv-19-2,ErIv-20}. For static case ($\eps = -1$) possible  applications of the obtained solutions may  be a subject of a further research, aimed at a search of topological black hole solutions (with flat horizon) or wormhole solutions which are coinciding  asymptotically  (for ($u \to \pm \infty$))  with our solutions.

 {\bf Acknowledgments}

This paper has been supported by the RUDN University Strategic Academic Leadership Program (recipients: V.D.I. - mathematical model development and  K.K.E. - simulation model development).  The reported study was (partially) funded by RFBR, project number 19-02-00346 (recipients  K.K.E. and V.D.I. - physical model development).

\end{document}